\documentclass[submission, Phys]{SciPost}
\hypersetup{
    colorlinks,
    linkcolor={red!50!black},
    citecolor={blue!50!black},
    urlcolor={blue!80!black}
}

\usepackage[bitstream-charter]{mathdesign}
\DeclareSymbolFont{usualmathcal}{OMS}{cmsy}{m}{n}
\DeclareSymbolFontAlphabet{\mathcal}{usualmathcal}

\begin{document}

\begin{center}{\Large \textbf{
The Single-Photon Annihilation of Channeled Positron in Si Crystal\\
}}\end{center}

% TODO: write the author list here. Use first name (+ other initials) + surname format.
% Separate subsequent authors by a comma, omit comma and use "and" for the last author.
% Mark the corresponding author with a superscript star.
\begin{center}
K.B.~Korotchenko\textsuperscript{1$\star$},
Y.P.~Kunashenko\textsuperscript{2}
\end{center}

\begin{center}
{\bf 1} National Research Tomsk Polytechnic University, Lenin Ave. 30, 634050 Tomsk, Russia
\\
{\bf 2} Tomsk State Pedagogical University, Kievskaya Str. 60, 634061 Tomsk, Russia
\\
${}^\star$ {\small \sf CorrespondingAuthor@korotchenko@tpu.ru}
\end{center}

\begin{center}
\today
\end{center}

% For convenience during refereeing (optional),
% you can turn on line numbers by uncommenting the next line:
%\linenumbers

\section*{Abstract}
{\bf

 The differential cross-section of the single-photon annihilation of channeled positron on the K-shell electron of one of the crystal atoms is calculated and angular distribution of emitted photons is investigated. It is shown that cross-section has a complicated dependence on positron entrance angle with respect to crystal plane. It is demonstrated that this orientation dependence is different for different positron transverse energy levels. The dependence of the differential cross-section on the electron longitudinal energy is studied.
}

\vspace{10pt}
\noindent\rule{\textwidth}{1pt}
\tableofcontents\thispagestyle{fancy}
\noindent\rule{\textwidth}{1pt}
\vspace{10pt}

\section{Introduction}
\label{sec:intro}

 It is well known that due to momentum and energy conservation laws free electrons and positrons can annihilate only into two or more photons. The situation changes if this process occurs in an external electromagnetic field when part of the transmitted momentum is taken by the field  and as a result, single-photon annihilation (SPA) becomes possible \cite{Akhiezer, Berestetskii}. At present, the phenomenon of SPA of the electron and the positron in the external fields was studied theoretically and experimentally.

 Fermi and Uhlenbeck \cite{Fermi} first considered the SPA of the free positron with the atomic electron. The more detailed theory was developed in \cite{Hulme,Nishina,Bethe,Jager,MgVoy,Moroi,Johnson,Johnson1,Broda,Mikhailov}. The cross-section of the SPA of relativistic positron with $K$-shell electron of the atom was calculated in \cite{Hulme}, where also the interaction of the positron with the atom was taken into account. In the paper \cite{Jager} it was discussed SPA of positrons, where the positive electron (positron) was regarded as a hole in the negative energy electrons distribution, and the process of annihilation was described as an atomic electron jumping down into this hole with the emission of a single photon.

 The cross-sections for the SPA of longitudinally polarized positrons and for the photoelectric effect produced by longitudinally polarized photons were derived in \cite{MgVoy}. A photoelectric effect and SPA in hydrogen with large momentum transfer were studied, taking into account the recoil and anomalous magnetic moment of proton in \cite{Moroi}. The calculations of the cross-section for SPA of positrons by $K$-shell electrons in the Coulomb field of a nucleus were presented in \cite{Johnson}. In this paper, the numerical results were given for nuclear charges $Z = 73,74,78,79,82$, and $90$ and for positron energies from threshold to $1.75$ MeV. Angular distributions of radiation arising from the SPA of positrons with $K$-shell electrons were calculated in \cite{Johnson1}. Numerical calculations of the differential and total cross-sections of the SPA of positrons in an atomic field were presented in \cite{Broda} for the $K$ and $L$ shells for atomic nuclei with $Z = 47$ to $Z = 92$. The process of annihilation of a fast positron with a $K$-shell electron, accompanied by emission of a photon and a second $K$-shell electron was studied in \cite{Mikhailov}.

 The experimental investigations of the SPA of positron in the field of atom was performed in \cite{Sodickson,Palathingal}.

 The SPA of an electron-positron pair in the pulsed light field was investigated theoretically in \cite{Voroshilo}. The pulsed laser field was described by the plane circularly polarized electromagnetic wave. The analytical expressions for the probability of the studied process were derived.

 The one- and two-photon annihilation processes of the electron-positron pairs in an ultra-strong magnetic field at a middle relativistic regime were considered in \cite{Lewicka,Lewicka1}. In the paper \cite{Bander} it was studied the annihilation of free electrons and positrons in intense magnetic fields. It was shown when both the electron and the positron are in the lowest energy state the resulting gamma rays emerge predominantly transverse to the field direction; in the case where one of the particles is in an excited state, the radiation is predominantly along the field direction.

 The interaction of charged particles with oriented crystal for a long time is a perceptive field of research both theory and experiment. The passage of particles through crystals is accompanied by various physical phenomena, for example, coherent bremsstrahlung and coherent pair production, channeling radiation, the electron-positron pair creation, and others. These phenomena are described in detail in a number of monographs and original articles (see for example \cite{Mikelian,Baryshevsky,Baier,Akhiezer1, Kimball,Uberall,Lasukov,Lasukov1,Lasukov2, Nitta,Nitta1,Olsen,Dabagov,KoKu,KoKu2,Kunashenko} and references therein).

 When relativistic positrons pass through the crystal in the channeling regime, one- and two-photon annihilation processes are possible. For the first time, SPA of the channeled positron was theoretically considered in \cite{Haakenaasen}. In this paper, the orientation dependence of SPA on the positron incidence angle with respect to the crystal plane was qualitatively studied. The annihilation of the channeled in crystal positron was experimentally studied in \cite{Hau,Hunt} in order to investigate the features of the crystal lattice. From the experimental point of view, it was more convenient to use a two-photon process.

 In the present paper, for the first time we carried out an exact calculation of the differential cross-section of SPA of the planar channeled positron, and for the first time, it is detailed investigated the angular distribution of photons arising in SPA of the channeled positrons with the $K$-shell electron of one of the crystal atoms was calculated.

\section{Single-photon annihilation probability}
\label{L2}

 The channeled phenomenon arises when a particle moved in the continuous potential of the system of crystal planes (axes) in the bound states. The positron passing through a crystal in a planar channeled mode can collide with a $K$-shell electron of one of the crystal atoms and as a result, electron and positron could annihilate. Because both electron and positron travel in the external fields, therefore, there is a possible SPA process.

 The single-photon annihilation of the electron-positron pair is described by a first-order Feynman diagram (Fig.1).
\begin{figure}[h]
\centering\noindent
\includegraphics[width=6cm]{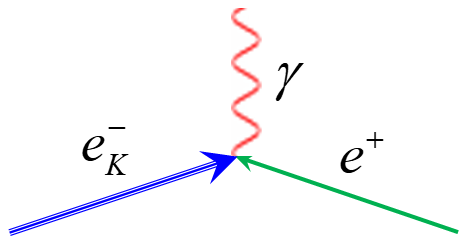}
\caption{Feynman diagram for single-photon annihilation of channeled positron on $K$-shell electron. The double arrow represents the $K$-shell electron of the atom, the single arrow is channeled positron and waved arrow is the photon.}\label{fig1}
\end{figure}

 In our case when the channeled positron annihilate with the electron of the atom of crystal this Feynman diagram coincides with one for positron Channeling Radiation (CR) if one replaced the wave function of the crystal atom $K$-shell electron by the positron wave function of the final state.

 Consequently, the probability of SPA of a channeled positron in a crystal is written similarly to the probability of radiation, but additionally, it is necessary to change the final state density of CR onto $d\varrho_f = L^3 d^3{\boldsymbol{\kappa}}/(2 \pi)^3$, where  and ${\boldsymbol{\kappa}}$ is the wave vector of the photon at the final states and $L$ is the normalization length
\begin{equation}
    dw_{if} = \frac{2 \pi}{\hbar} \mid M_{if}\mid^2 \delta(E_i - E_f) d\varrho_f\: . \label{eq1}
\end{equation}
 Here $M_{if}$ is a matrix element of SPA of channeled positron with the $K$-shell electron of one of the crystal atoms. According to Feynman diagram (Fig.1), the matrix element has a form
\begin{equation}
    M_{if} = - e \int j^\mu_{if} A^{\ast}_\mu d^3 {\bf r}\: , \label{eq2}
\end{equation}
 where the 4-vector $A_{\mu}$ represents the wave function of a photon, $j^\mu_{if} = \bar{\Psi}_f \boldsymbol{\gamma}^\mu \psi_i$ is the particles 'current operator', $\boldsymbol{\gamma}^\mu$ is the Dirac matrix, $\psi_i$ is the wave function of a channeled positron and $\Psi_f$ is the wave function of a $K$-shell electron.

 If for the 4-vector $A_\mu$ we chose Coulomb gauge ${\bf \nabla A} = 0$ and take into account $\varphi_0 = A_0 = 0$, then we obtain $j^\mu_{if} A^{\ast}_\mu \Rightarrow {\bf j}_{if} {\bf A}$ with
\begin{equation}
    {\bf A} = \sqrt{\frac{2 \pi \hbar c^2}{\omega L^3}} e^{\mathbf{i}\boldsymbol{\kappa}\mathbf{r}}\boldsymbol{\epsilon}_{\boldsymbol{\kappa}} \: , \label{eq3}
\end{equation}
 where $\boldsymbol{\epsilon}_{\boldsymbol{\kappa}}$ is the photon polarization vector.

 It is convenient to choose the coordinate axes as follows: the X-axis perpendicular to the crystal plane, the Z and Y axes parallel to the crystal plane, and the positron longitudinal momentum is directed along Z-axis. Then wave function of a planar channeled positron in the initial quantum state $i$ can be written in the form \cite{Beloshitsky}
\begin{eqnarray}
 \label{eq4}
 \psi_i(\mathbf{r}) & = & \sqrt{\big(E_{i \|} + mc^2\big)\big/E_{i\|}}u_i\phi_i(\mathbf{r}_{\perp}) e^{\mathbf{i}\mathbf{p}_{i\|}\mathbf{r}_{\|}/\hbar}\: , \nonumber\\
  u_i & = & \left(
\begin{array}{c}
  w \\
  \frac{\boldsymbol{\sigma}\cdot\hat{\mathbf{p}}c}{c^2 m + E_{i\|}}
\end{array}
  \right)\: .
\end{eqnarray}
 Here $w$ is the 2D-spinor normalized by the condition  $w^{+}w = 1$, $\boldsymbol{\sigma}$ are the Pauli matrixes. The transverse wave function $\phi_i(\mathbf{r}_{\perp})$ describes the quantum states of the relativistic channeled particle and satisfies the Schr\"odinger-like equation \cite{Kimball,Beloshitsky} with relativistic mass $\gamma m$ ($\gamma = E_{\|}/m c^2$ is Lorentz factor).

 As for the final state, for our purposes, the wave function of the electron in the atom can be chosen in a simple enough form
\begin{equation}
 \Psi_f = u_K\Psi_K \: , u_K =
 \left(\begin{array}{c}
  w \\
  0
\end{array}
  \right)\: . \label{eq06}
\end{equation}
 Here $\Psi_K$ is a nonrelativistic $K$-shell electron wave function (solution of the Schr\"odinger equation, see below), $w$ is again the 2D-spinor.

 Substitution of defined in such a way wave functions into the expression of 'current operator', after some algebra, results in the following
\begin{equation}
 {\bf j}_{if} {\bf A} =
 {C_M}\Psi_K^{\ast} (w^{+}(c\boldsymbol{\sigma}\cdot\hat{\mathbf{p}})
 (\boldsymbol{\sigma}\cdot\boldsymbol{\epsilon}^{\ast}) w)\phi_i(x) e^{-\mathbf{i}\boldsymbol{\kappa}\mathbf{r}} e^{\mathbf{i}\mathbf{p}_{i\|}\mathbf{r}_{\|}/\hbar}\: , \label{eq5}
\end{equation}
 where we introduce the following definition
\begin{equation}
    C_M = \frac{1}{L} \sqrt{\frac{E_{i\|} + mc^2}{E_{i\|}}}\sqrt{\frac{2\pi\hbar c^2}{\omega L^3}} \frac{1}{E_{i\|} + mc^2} \: . \label{eq6}
\end{equation}

 Usually, the initial beam of channeled particles is not polarized. Therefore, formula (\ref{eq5}) should be averaged over the spin states of the positron. In addition, it is necessary to average over the spin states of the electron in the crystal atom. After that we have
\begin{equation}
     {\bf j}_{if} {\bf A} = {C_M}\Psi_K^{\ast} c(\hat{\mathbf{p}} \cdot\boldsymbol{\epsilon}^{\ast})\phi_i(x) e^{-\mathbf{i}\boldsymbol{\kappa}\mathbf{r}} e^{\mathbf{i}\mathbf{p}_{i\|}\mathbf{r}_{\|}/\hbar} \: , \label{eq7}
\end{equation}

 Now the matrix element $M_{if}$ takes the form
\begin{equation}
    M_{if} = \boldsymbol{\mathfrak{L}}\cdot\boldsymbol{\epsilon}^{\ast}\: , \label{eq8}
\end{equation}
 where we denote
\begin{equation}
     \boldsymbol{\mathfrak{L}} = -e c C_M\int\Psi_K^{\ast}e^{-\mathbf{i}\boldsymbol{\kappa}\mathbf{r}} \hat{\mathbf{p}}\: \phi_i(x) e^{\mathbf{i}\mathbf{p}_{i\|}\mathbf{r}_{\|}/\hbar}d^3 {\bf r} \: , \label{eq9}
\end{equation}

\section{Calculation of the single-photon annihilation matrix element}
\label{L3}

 First, let us consider the transverse wave function $\phi_i(x)$ of the planar channeled positron. Due to crystal periodicity in a direction perpendicular to crystal planes, this function should be the Bloch one and satisfy the Schr\"{o}dinger-like equation.

 The method of the solution of this equation was developed in \cite{Ashcroft,Kittel} and is well known now. According to \cite{Ashcroft,Kittel} transverse wave function $\phi_i(x)$ has the following form
\begin{equation}
    \phi_i(x) = \sum_{m} e^{\mathbf{i} x G_{m,i_n}} X_{i,i_n,m}\: , \label{eq10}
\end{equation}
 where $X_{i,i_n,m}$ is Fourier components (with number $m$) of the wave function, $G_{m,i_n} = 2\pi  m/d$ $+ \pi i_n/10 d$, $d$ is an interplanar distance.

 To take into account that the energy levels of a particle in a periodic potential are energy bands, we divide each band into ten subbands $n = 10$ and introduced the following index $i_n$ for numbering $n$-th subband of the band $i$.

 Substitution of the wave function (\ref{eq10}) into vector $\boldsymbol{\mathfrak{L}}$ results in:
\begin{eqnarray}
 \hspace{-10mm} & & \boldsymbol{\mathfrak{L}}_x = -e c C_M \hbar G_{m,i_n} \int\Psi_K^{\ast}e^{-\mathbf{i}\boldsymbol{\kappa}\mathbf{r}} \phi_i(x)e^{\mathbf{i}\mathbf{p}_{i\|}\mathbf{r}_{\|}/\hbar}d^3 {\bf r}\: , \label{eq11}\\ \hspace{-10mm} & & \boldsymbol{\mathfrak{L}}_y = -e c C_M p_y
 \int\Psi_K^{\ast}e^{-\mathbf{i}\boldsymbol{\kappa}\mathbf{r}} \phi_i(x) e^{\mathbf{i}\mathbf{p}_{i\|}\mathbf{r}_{\|}/\hbar}d^3 {\bf r}\: , \label{eq12}\\ \hspace{-10mm} & & \boldsymbol{\mathfrak{L}}_z = -e c C_M p_z
 \int\Psi_K^{\ast}e^{-\mathbf{i}\boldsymbol{\kappa}\mathbf{r}} \phi_i(x) e^{\mathbf{i}\mathbf{p}_{i\|}\mathbf{r}_{\|}/\hbar}d^3 {\bf r}\: . \label{eq13}
\end{eqnarray}

 Let us consider in more detail the calculation of the $\boldsymbol{\mathfrak{L}}_x$ component. Substitute positron transverse wave function (\ref{eq10}) into (\ref{eq11}) we  rewrite $\boldsymbol{\mathfrak{L}}_x$ in the following form
\begin{equation}
    \boldsymbol{\mathfrak{L}}_x = e c C_M \sum_{m_i} Ix_{i_n,m_i} X_{i,i_n,m_i}\: . \label{eq14}
\end{equation}
 where
\begin{equation}
 Ix_{i_n,m_i} = \hbar G_{m,i_n}\int e^{-\mathbf{i}\kappa_x x}e^{\mathbf{i}G_{m,i_n} x)}
 \left(\int \Psi_K^{\ast}(x,y,z) e^{\mathbf{i}(p_y/\hbar - ky) y}e^{\mathbf{i}(p_z/\hbar - kz)z}dy dz\right)dx \label{eq15}
\end{equation}

 The wave function of the electron of crystal atom $K$-shell can be written as the Slater wave function \cite{Bunge}.
 For electron on the $K$-shell the Slater wave function \cite{Bunge} does not depend on angle variables, i.e. $\Psi_K(r,\theta,\phi) \rightarrow \Psi_K(r)$. For the further calculation the function $\Psi_K(r)$ is borrowed from the paper \cite{Bunge}:
\begin{equation}
    \Psi_K(r) = \sum_{p}\frac{(2\zeta_{0,p})^{3/2} e^{-\zeta_{0,p}r}(2\zeta_{0,p}r)^{{n\lambda}_{0,p}-1}C_{p,n,0}}
    {\sqrt{\left(2 n\lambda_{0,p}\right)!}} \: . \label{eq16}
\end{equation}
 Here $C_{p,n,0}$, ${n\lambda}_{0,p}$ and $\zeta_{0,p}$ are parameters defined in \cite{Bunge}.

 Taking into account the identity
\begin{equation}
 \frac{(2\zeta_{0,p}r)^{n\lambda_{0,p}-1}}{e^{\zeta _{0,p}r}} = (-2\zeta_{0,p})^{n\lambda_{0,p}-1}(\partial_{\zeta_{0,p}}^{{n\lambda}_{0,p}-1} e^{-\zeta_{0,p}r})\: . \label{eq17}
\end{equation}
 we rewrite the integral over $dy dz$ as follows
\begin{eqnarray}
 & & \hspace{-10mm} \int ... dy dz = \sum_{p}\frac{(2\zeta_{0,p})^{3/2}C_{p,n,0}}
    {\sqrt{\left(2 n\lambda_{0,p}\right)!}}
    (-2\zeta_{0,p})^{{n\lambda}_{0,p}-1} \label{eq18} \\
 & & \hspace{-5mm} \partial_{\zeta_{0,p}}^{{n\lambda}_{0,p}-1} \int e^{-\zeta_{0,p}r} e^{\mathbf{i}(p_y/\hbar - ky)y}e^{\mathbf{i}(p_z/\hbar - kz)z}dy dz\: . \nonumber
\end{eqnarray}

 To calculate the new integral in (\ref{eq18}) we use a cylindrical coordinate system ($\rho, \varphi$). The integration over angle $\varphi$ gives
\begin{eqnarray}
 & & \int e^{-\zeta_{0,p}r} ... dy dz = \int e^{-\zeta_{0,p}\sqrt{\rho^2+x^2}} ... \rho d\rho d\varphi = \nonumber\\
 & & 2\pi \int e^{-\zeta_{0,p}\sqrt{\rho^2+x^2}} J_0(\rho  \sqrt{q_y^2+q_z^2})\rho d\rho\: . \label{eq19}
\end{eqnarray}
 Here $J_0(z)$ is the Bessel function, $q_y = p_y/\hbar-k_y$, $q_z = p_z/\hbar-k_z$. As a result, we get the tabular integral \cite{Prudnikov}
\begin{eqnarray}
 & & \hspace{-10mm} I_1 = \int e^{-\zeta_{0,p}} ... \rho d\rho = \zeta_{0,p} (\zeta_{0,p}^2+q_y^2+q_z^2)^{-3/2} \times\nonumber\\
 & & \hspace{-10mm} \times \left(x\sqrt{\zeta_{0,p}^2 + q_y^2 + q_z^2}+1\right) e^{-x \sqrt{\zeta_{0,p}^2 + q_y^2+q_z^2}}\: . \label{eq20}
\end{eqnarray}

 Finally, we arrive at the following expression for the (\ref{eq18})
\begin{equation}
 \int ... dy dz = 2\pi\sum_{p}\frac{(2\zeta_{0,p})^{3/2}C_{p,n,0}}
    {\sqrt{\left(2 n\lambda_{0,p}\right)!}}
    (-2\zeta_{0,p})^{{n\lambda}_{0,p}-1}
 \partial_{\zeta_{0,p}}^{{n\lambda}_{0,p}-1} I_1\: . \label{eq21}
\end{equation}

 Similar transformations can be done for the rest components $\boldsymbol{\mathfrak{L}}_y$ and $\boldsymbol{\mathfrak{L}}_z$. For this, generalizing formula (\ref{eq17}), we introduce the function
\begin{equation}
 DI_{p,q_{yz}}(x) = \frac{\partial^p}{\partial \zeta_{0,p}^p}\frac{\zeta_{0,p}\Big(x\sqrt{q_{yz}^2+\zeta_{0,p}^2}+1\Big)
 e^{-x\sqrt{q_{yz}^2+\zeta_{0,p}^2}}}{(q_{yz}^2+\zeta_{0,p}^2)^{3/2}} \: , \label{eq22}
\end{equation}
 where $q_{yz}^2 = q_y^2+q_z^2$. Now we obtain
\begin{eqnarray}
 \boldsymbol{\mathfrak{L}}_x & = & -e c C_M \hbar \sum_m G_{m,i_n} X_{i,i_n,m} Io_{m,i_n}\: , \label{eq23}\\
 \boldsymbol{\mathfrak{L}}_y & = & -e c C_M p_y\sum_m X_{i,i_n,m} Io_{m,i_n}\: , \label{eq24}\\
 \boldsymbol{\mathfrak{L}}_z & = & -e c C_M p_z\sum_m X_{i,i_n,m} Io_{m,i_n}\: , \label{eq25}
\end{eqnarray}
 where we introduce the following notation:
\begin{equation}
 Io_{m,i_n} = 2\pi\sum_p \frac{(2\zeta_{0,p})^{3/2} C_{p,n,0}} {\sqrt{2 n\lambda}_{0,p}!}
 (-2\zeta_{0,p})^{{n\lambda}_{0,p}-1}IDI_{p,q_{yz}}\: . \label{eq26}
\end{equation}
 Here the notation $IDI_{p,q_{yz}}$ is the integrals of the derivatives (\ref{eq22}) $DI_{p,q_{yz}}(x)$
\begin{equation}
    IDI_{p,q_{yz}} = \int_0^d DI_{p,q_{yz}}(x)dx\: . \label{eq27}
\end{equation}

 After substitution (\ref{eq22}) into (\ref{eq27}) those integrals can be easily analytically calculated. The results depend on index $p$ (from $1$ to $10$) therefore, it is necessary to calculate $10$ integrals. So for example for $p < 3$ we have
 \begin{eqnarray}
    IDI_{p,q_{yz}} & = & \frac{e^{-q_z\zeta_p d}\zeta_{0,p}}{q_z\zeta_p^3 (q_x^2+q_z\zeta_p^2)^2}\Big(q_x [q_x^2 (q_z \zeta_p d + 1)+q_z\zeta_p^2(q_z \zeta_p d + 3)] \sin(q_x d) - \nonumber\\
    & - & q_z\zeta_p^2[(q_x^2+q_z\zeta_p^2)d+2 q_z\zeta_p] \cos(q_x d)+2 q_z\zeta_p^3 e^{q_z \zeta_p d}\Big)\: . \label{eq271}
\end{eqnarray}

 For other values of the parameter $p$, we have similar expressions but are longer, cumbersome, and difficult for visual perception. Therefore, we do not give the rest formulas here.

 As we don't interest photon polarization we sum up matrix element (\ref{eq8}) squared over the photon polarizations by the well-known formula \cite{Akhiezer, Berestetskii}
\begin{equation}
 \overline{(\boldsymbol{\mathfrak{L}}\cdot\boldsymbol{\epsilon}^{\ast})
 (\boldsymbol{\mathfrak{L}}^{\ast}\cdot\boldsymbol{\epsilon})} = (\boldsymbol{\mathfrak{L}}\times\boldsymbol{n})\cdot
 (\boldsymbol{\mathfrak{L}}^{\ast}\times\boldsymbol{n})\: , \label{eq26}
\end{equation}
 where ${\bf n} = (\sin\Theta\cos\Phi,\sin\Theta\sin\Phi,\cos\Theta)$ is a photon emission direction.

 Finally, the matrix element $M_{if,i_n}$ squared is written as
\begin{eqnarray}
 \mid M_{if,i_n}\mid^2 & = & (c C_M\hbar)^2 \Bigg\{\Bigg|\cos\Theta \sum_{m_i}G_{m,i_n} X_{i,i_n,m} Io_{m,i_n} - \frac{p_z}{\hbar} \cos\Phi\sin\Theta \sum_{m_i} X_{i,i_n,m} Io_{m,i_n}\Bigg|^2 + \nonumber\\
 & + & \Bigg| \sum_{m_i}G_{m,i_n} X_{i,i_n,m} Io_{m,i_n} \Bigg|^2 + \Bigg(\frac{p_z}{\hbar}\sin\Theta\sin\Phi\Bigg)^2 \Bigg|\sum_{m_i} X_{i,i_n,m} Io_{m,i_n}\Bigg|^2 \Bigg\}\: . \label{eq28}
\end{eqnarray}

\section{Numerical results}
\label{L4}

 Substituting the obtained matrix element (\ref{eq28}) into (\ref{eq1}) and dividing the result by the flux of initial particles $J$, we find the cross-section of the SPA process $d\sigma = dw_{if,i_n}/J$
\begin{equation}
    d\sigma = \langle P_{\theta,i_n} dw_{if}\rangle_{i_n}/J\: . \label{eq29}
\end{equation}

 Here $J = v/V$ with $v$ being positron velocity and $V = L^2 d$ being normalization volume, where $L$ is normalization length and $d$ is the distance between neighbor crystal planes. In the equation (\ref{eq29}) notation $\langle...\rangle_{i_n}$ means an averaging over $i$-th energy band. The physical meaning of the indices $i_n$ is described above (before formulas (\ref{eq11}-\ref{eq13})). Additionally, during averaging, we take into consideration an initial population of the $n$-th section $P_{\theta,i_n}$ of the $i$-th energy band of the positron's transverse motion.  The initial population $P_{\theta,i_n}$ depends on the entry angle $\theta$ relative to the crystal plane.

 The continuous approximation for the crystal planes system potential is valid only for the relativistic particles, therefore, we have to limit the minimal positron energy. We choose a silicon crystal, oriented by the $(110)$ planes as a target, and perform numerical calculations for the positron energy range $50 - 100$ MeV. The obtained results are as follows.

\begin{figure}[h]
\centering\noindent
\includegraphics[width=55mm]{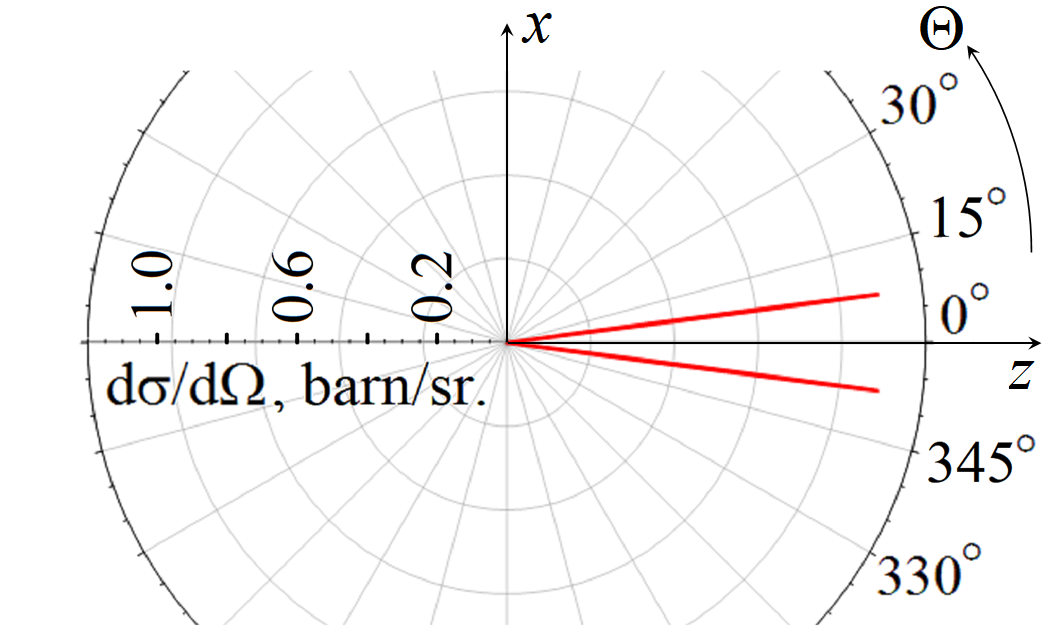}
\caption{Polar plot of the angular distribution of emitted photons (red lines) during SPA of a channeled positron with an energy of $60$ MeV on the K-electron of the atom of Si crystal. The red lines length denotes the value of the cross-section. The $x$ and $z$ are the cartesian coordinate axes. The positron entry angle with respect to the crystal plane is equal to $\theta \simeq 0.5 \theta_C$. The calculation was performed for the annihilation of channeled positrons from the first ($i = 1$) excited energy level.}\label{fig2}
\end{figure}

 Fig.~\ref{fig2} shows the result of calculating the angular distribution of photons arising due to the SPA of positrons with $60$ MeV energy channeled along the $(110)$ planes of the Si crystal. The positron initial angle is equal to $\theta = 0.5 \theta_C$ with respect to the crystal plane, where $\theta_C$ is the critical channeling angle (Lindhard angle).  Fig.~\ref{fig2} is plotted in polar coordinates.

 From Fig.\ref{fig2}, it can be seen, photon arising as a result of SPA of a channeled $60$ MeV positron at the K-electron of the atom of Si crystal is emitted at an angle $\Theta$ of the order of $\sim 6.2^{\circ}$ to the positron direction of motion (axis z). In this case, the differential cross-section (over solid angle $d\Omega$) $d\sigma/d\Omega$ is of the order of $\sim 1$~barn/sr.
\begin{figure}[h]
\centering\noindent
\includegraphics[width=35mm]{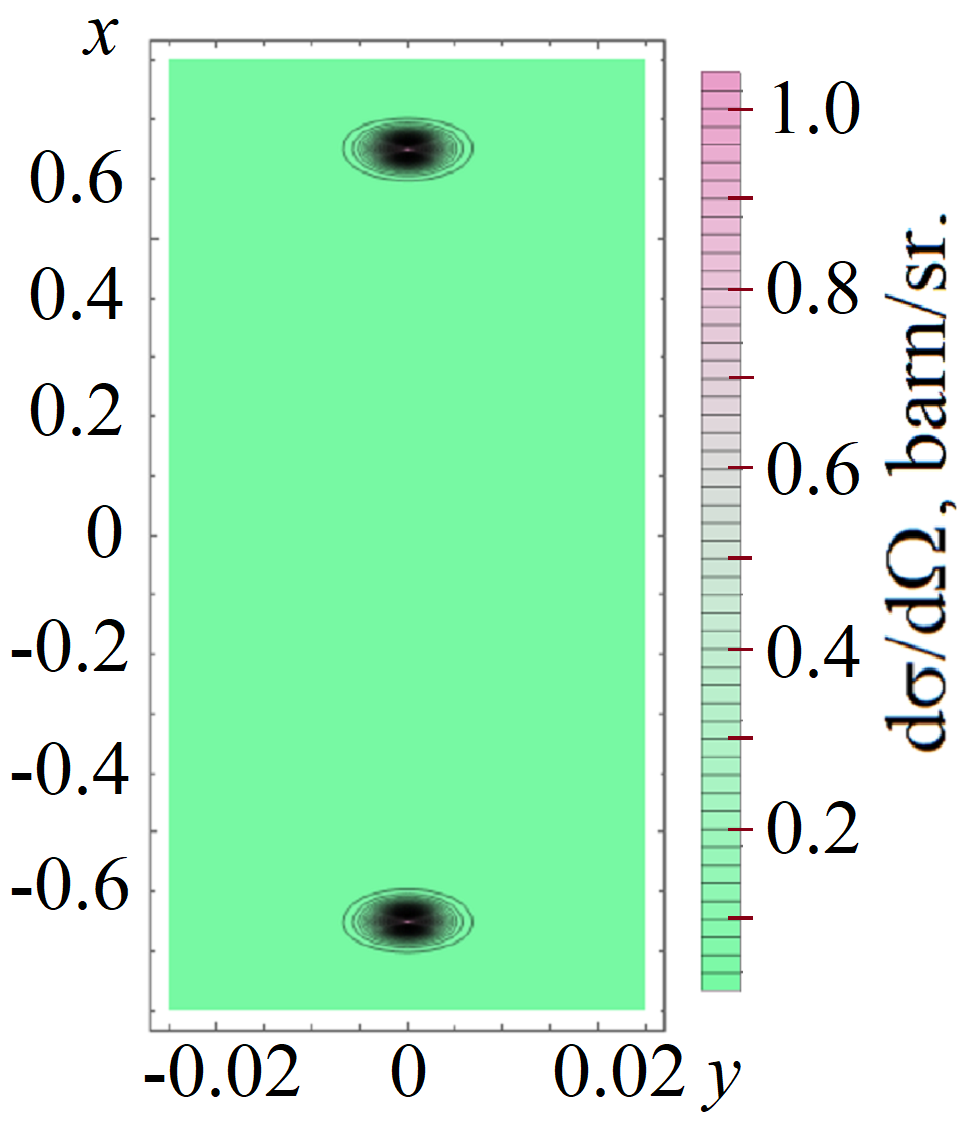}
\caption{Contour plot of the photon angular distribution of SPA of a channeled positron. The parameters are as in Fig.\ref{fig2}.}\label{fig3}
\end{figure}

 To clarity, Fig.3 shows the same result, but as a contour plot. The $x$ and $y$ coordinates are in relative units (the $x$-axis is perpendicular to the channeling planes). At some space point (for example, $z = 0$) the positron annihilates and emits a photon that moves at a small angle to the Z-axis. This photon collides on an imaginary screen placed at a distance of $5$ 'arbitrary' units, the $x$ and $y$ are coordinates of the photon on this screen. The plane of the screen is perpendicular to the Z-axis (to longitudinal positron momentum $p_z$). The figure shows, the picture that can be observed experimentally on this screen. Different colors correspond to different values $d\sigma/d\Omega$.

\begin{figure}[h]
\centering\noindent
\includegraphics[width=6cm]{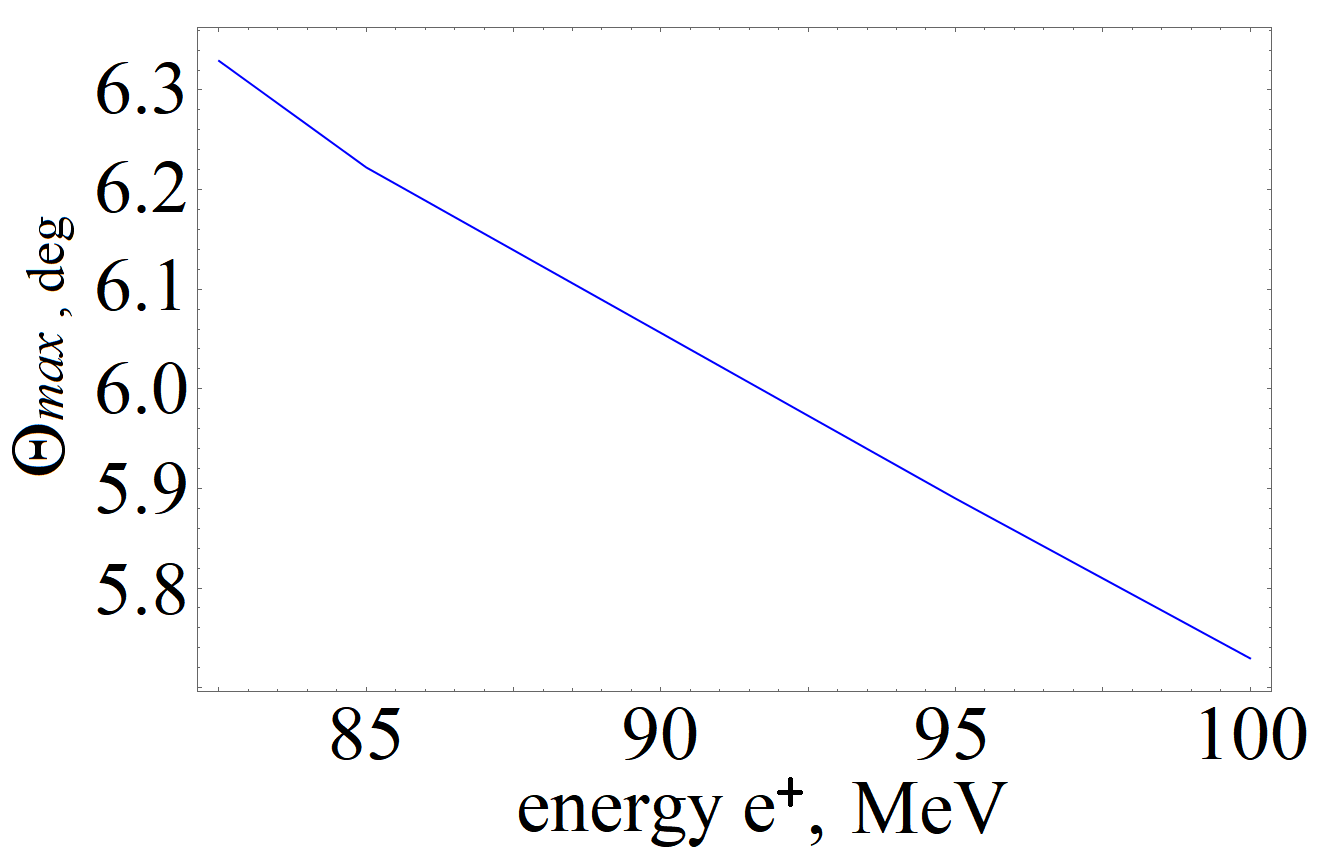}
\caption{Dependence of the angle $\Theta_{max}$ on the energy of the channeled positron. The $\Phi$ angle is fixed and equals $\Phi = 0$.}\label{fig4}
\end{figure}

 Let us introduce the angle of photon emission $\Theta_{max}$ at which the differential cross-section of the positron SPA on $K$-shell electron reaches its maximum value $d\sigma_{max}/d\Omega$. It is obvious that $\Theta_{max}$ changes with changing positron energy. We investigated this dependence by formulas (\ref{eq28}, \ref{eq29}) obtained above. The result of this study is present in Fig.~\ref{fig4}, where we plot the angle $\Theta_{max}$ value versus positron energy. The polar angle $\Phi$ is fixed and equals $0^{\circ}$.

 It is very important to notice that the angle of emission of photons $\Theta_{max}$ corresponding to the maximum of the differential cross-section $d\sigma_{max}/d\Omega$ is independent on either angle $\theta$ of entry of positrons into the crystal or sub-barrier energy level $i$ from which the positron annihilates.

\begin{figure}[h]
\centering\noindent
\includegraphics[width=75mm]{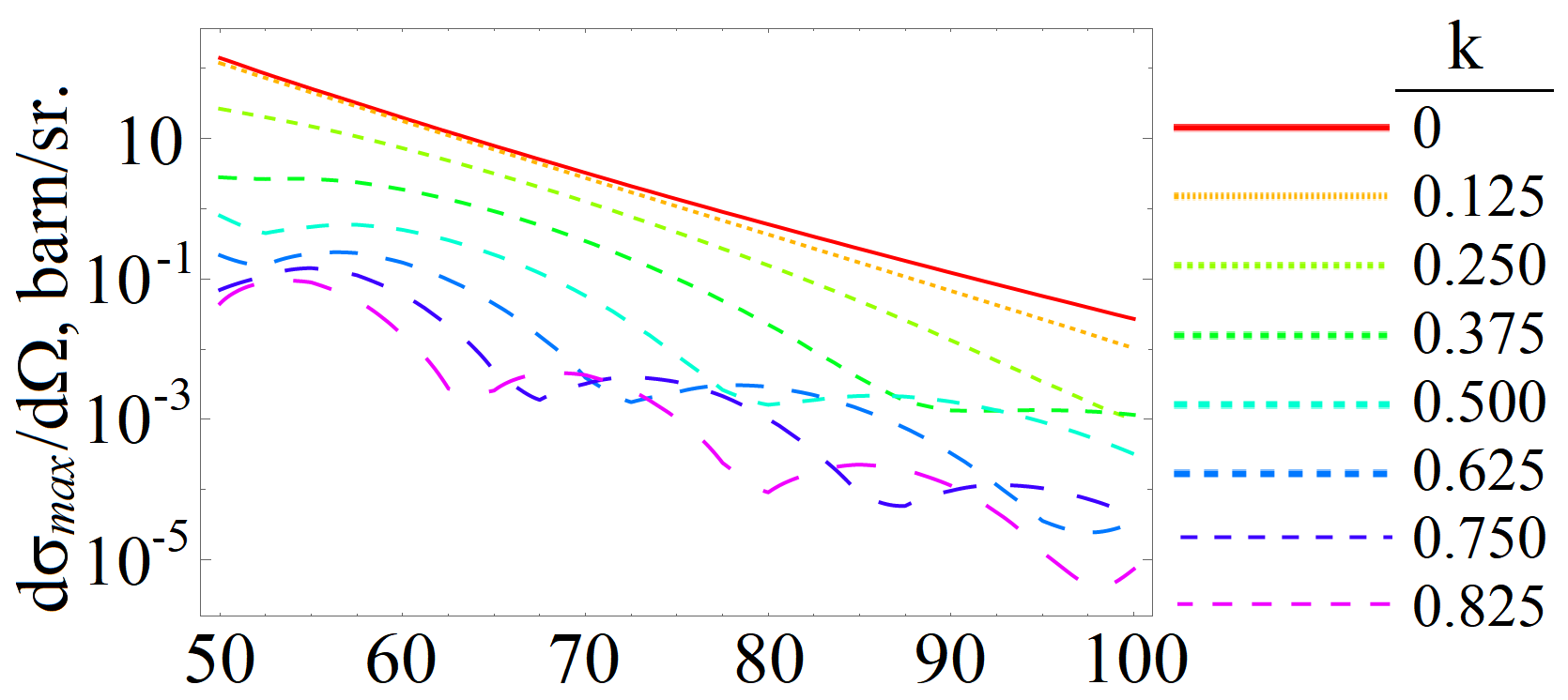}
\caption{Dependencies of the maximum value of differential cross-section $d\sigma_{max}/d\Omega$ on the longitudinal energy of the channeled positron. The number $k$ corresponds to different angles of entry of the positron into the crystal $\theta = k \theta_C$ relative to the channeling plane. The calculation is performed for the annihilation of channeled positrons from the first ($i = 1$) excited transverse energy level. The ordinate uses a logarithmic scale.}\label{fig5}
\end{figure}

 In Fig.~\ref{fig5} it is shown the maximal value $d\sigma_{max}/d\Omega$ of the differential cross-section $d\sigma/d\Omega$ of SPA of positrons channeled in Si crystal as a function of its energy for 8 different angles of incidence of the positron (from $\theta = 0$ to $\theta = 0.825\theta_C$). The calculation was performed for the annihilation of channeled positrons from the first ($i = 1$) excited transverse energy level.

 Consider Fig.~\ref{fig5} in more detail. For $k = 0$ (positron entry angle $\theta = 0$) one can see that curve is a nearly straight line. A logarithmic scale is used on the ordinate, therefore, the dependence of the maximal value of the differential cross-section $d\sigma_{max}/d\Omega$ from the energy of channeled positron describes by exponential law. We fit this dependence by function
\begin{equation}
    d\sigma_{max} = \sigma_0~e^{-\eta\gamma}\: . \label{eq30}
\end{equation}
 where $\sigma_0 \simeq 5.6\times10^5$ barn/sr, $\eta = -0.085$. The relative error does not exceed $8\%$.

 As for other curves, we see their oscillations with increasing the longitudinal energy of the channeled positron. The larger the entry angle, the shorter the period of the oscillations and the greater their amplitude. These oscillations occur against the background of a decrease in the maximal value of the differential cross-section with an increase in energy approximately exponentially.

 In addition, it is important to note that, similar to Parametric X-Radiation at the Channeling (PXRC), the breakpoints of the curves correspond to the value of positron longitudinal energy at which a new odd level (band) appears, for under-barrier energy \cite{KKB-PXR}.

\section{Discussion}
\label{L5}

 As far as we know there is only one work devoted to the theoretical consideration of SPA of a channeled positron on the K-electron of an atom of crystal \cite{Kalashnikov}.

 In the paper \cite{Kalashnikov}, based on a qualitative analysis of the populations of transverse quantum states of a channeled positron it was studied the dependence of the cross-section on the angle between positron momentum and crystal plane. It was concluded for a positron with energy $25$ MeV channeled in (110) Si that with increasing of the entry angle of the positron into the crystal the differential cross-section increase.

\begin{figure}[h]
\centering\noindent
\includegraphics[width=130mm]{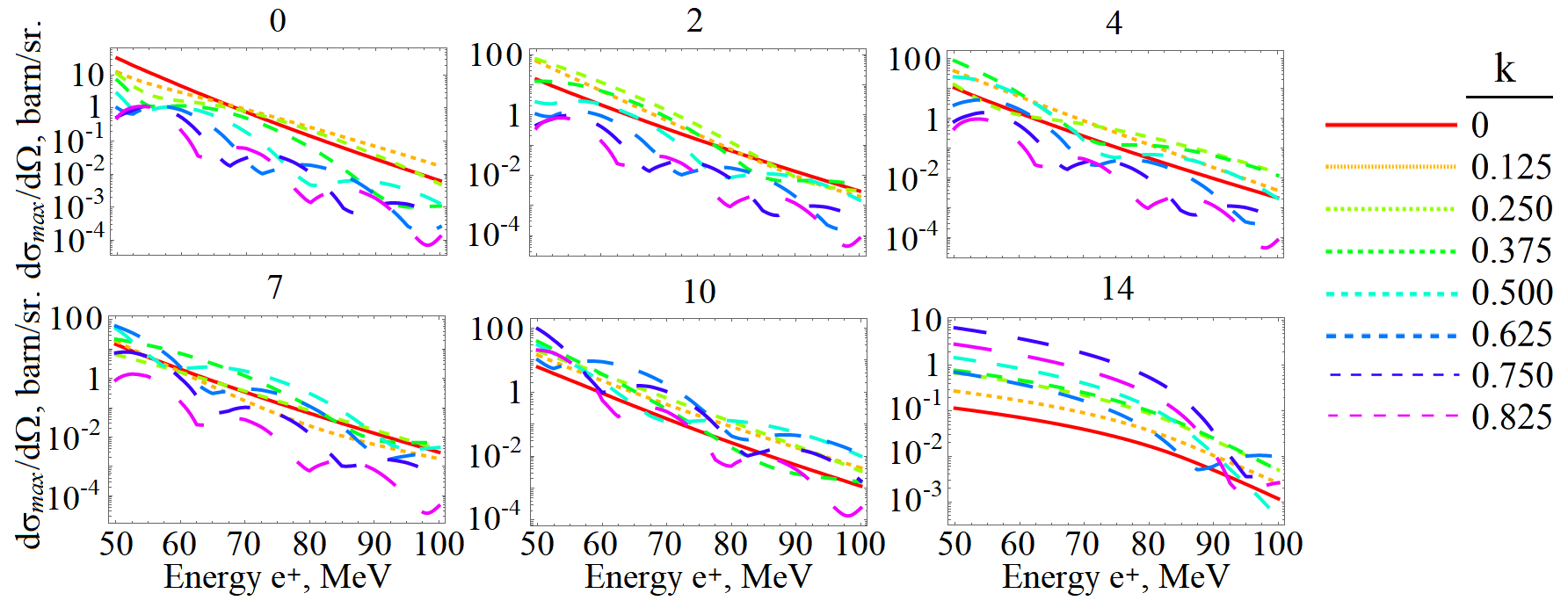}
\caption{Dependences of the maximum value $d\sigma_{max}/d\Omega$ of the differential cross-section $d\sigma/d\Omega$ on the longitudinal energy of channeled positrons entering into the crystal at different angles $\theta = k \theta_C$ to the channeling plane for different transverse energy bands (similar to Fig.~\ref{fig5} notation is used). The numbers $0...14$ on the picture are the numbers of the sub-barrier energy levels of the channeled positron. The ordinate uses a logarithmic scale.}\label{fig6}
\end{figure}

 Our calculations show that for the positrons channeled in the same crystal, the maximum of the cross-section $d\sigma_{max}/d\Omega$ of SPA behaves much more complicated (see Fig.~\ref{fig5}, \ref{fig6}).

 Let us consider the behavior of the maximum value $d\sigma_{max}/d\Omega$ of the differential cross-section $d\sigma/d\Omega$ on the angle of entry of the positron into the crystal for different sub-barrier bands of transverse energy (Fig.~\ref{fig5}, \ref{fig6}). For positrons that occupy low levels of transverse energy, the following tendency is observed: the value $d\sigma_{max}/d\Omega$ decreases with an increasing angle of entry of the positron into the crystal. This is most clearly manifested for positrons of the first excited transverse level $i = 1$ (see Fig.~\ref{fig5}). With an increase in the number of the energy band of the transverse energy of the positron, the picture changes. The magnitude of the maximum value $d\sigma_{max}/d\Omega$ for large angles of entry of the positron into the crystal gradually increases, and for small angles of entry, it decreases. Finally, for positrons located at the highest sub-barrier transverse energy band, the picture becomes the opposite: the magnitude of the maximum value $d\sigma_{max}/d\Omega$ of the differential cross-section $d\sigma/d\Omega$ strictly increases with increasing angle of entry of the positron into the crystal. Moreover, the values $d\sigma_{max}/d\Omega$ itself are different for different transverse energy bands. All the calculations are performed taking into account the populations of different transverse energy levels.

 In a real experiment, at any angle of incidence (not exceeding the Lindhard angle) of the positron into the crystal, all sub-barrier levels of the transverse energy are populated (with some probability) and, therefore, all curves in Fig.~\ref{fig5} and \ref{fig6} ($k = 0...14$) must be summed up and the pictures described above is not available for observation. Instead, some average curve will be observed.

 Neve the less, in the case of SPA of a channeled positron on the K-electron of an atom of crystal, some kind of orientation dependence should be observed. A detailed study of orientation dependence is planned to be carried out in a separate work. Overall, the results of our calculations confirm the qualitative conclusion of \cite{Kalashnikov} about the orientation dependence of the process under study.

 From Fig.~\ref{fig5} - \ref{fig6} follows that for all incident positron angles and for all positron energies bands the maximum value $d\sigma_{max}/d\Omega$ of the differential cross-section $d\sigma/d\Omega$ decreases with increasing positron energy (i.e. relativistic factor).

 The obtained results should be compared with ones for positron annihilation with $K$-shell electron of a separate atom.

 The cross-section of SPA of positron with $K$-shell electron of a separate hydrogen-like atom with the different atomic number $Z$ was studied in \cite{Fermi,Hulme,Nishina,Bethe,Jager,MgVoy,Moroi,Johnson,Johnson1,Broda}. The various approximations for the wave function of $K$-shell electron and for positron one in the electric field of the atom were used in these papers.

 In order to have an idea about the dependence of SPA of positron with $K$-shell electron of a separate atom on positron energy we use the simplest expression \cite{Akhiezer, Berestetskii}, which for our purpose it is convenient to rewrite in a next form
\begin{eqnarray}
 & &\hspace{-10mm} \frac{d\sigma^1}{d\Omega} = \frac{\alpha^6 Z^5 \hbar^2}{2 c^2 m^2}(\gamma^2-1)(4+\Big(\beta\gamma (1 + \gamma) \cos\Theta - \nonumber \\
 & &\hspace{-10mm} - \gamma(1+\gamma)\Big)(3+\gamma)/\gamma^3(1 + \gamma)^5 (1-\beta\cos\Theta^4)
 \label{eq30}\: .
\end{eqnarray}
 where $\beta = v/c$ and $\gamma$ is the relativistic factor $\gamma = (1 - \beta^2)^{-1/2}$, $v$ is the positron longitudinal velocity, $\alpha$ is fine-structure constant.

\begin{figure}[h]
\centering\noindent
\includegraphics[width=55mm]{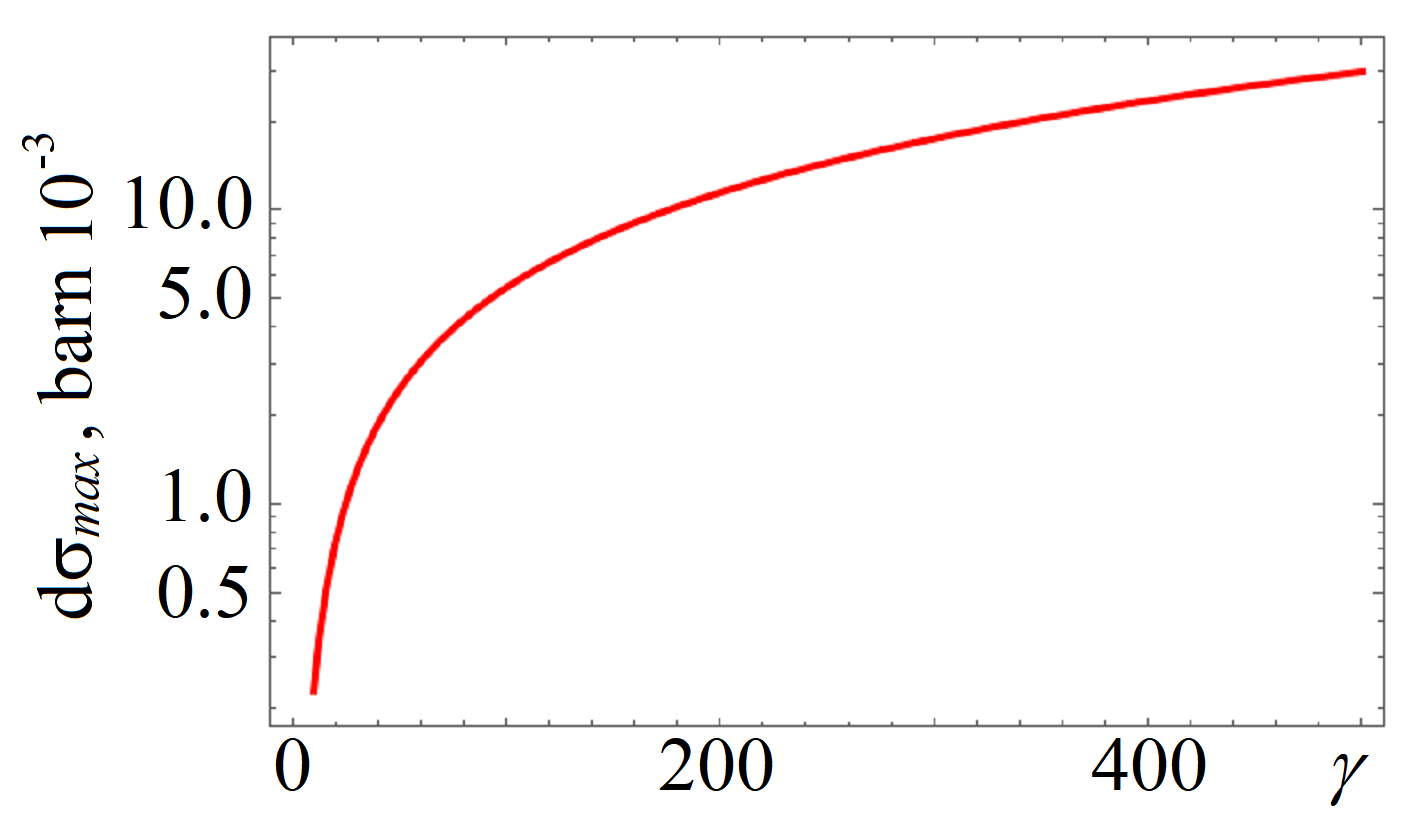}
\caption{Dependences of the maximum value of the cross-section $d\sigma^1_{max}$ on the relativistic factor of channeled positrons (a logarithmic scale is used).  We had chosen the $Z = 14$, i.e. Si atom.}\label{fig8}
\end{figure}

 Using the formula (\ref{eq30}) it is easy to find the angle of photon emission $\theta_{max}$ at which the differential cross-section of the positron SPA on $K$-shell electron reaches its maximum value $d\sigma^1_{max}/d\Omega$. Then one can calculate the maximum value of differential cross-section $d\sigma^1_{max}/d\Omega$ as a function of the positron relativistic factor. The calculation result is shown in Fig.~\ref{fig8}.

 One can see, Fig.~\ref{fig8} is observed opposite picture - the maximum magnitude of the differential cross-section $d\sigma^1_{max}/d\Omega$ increasing with increasing positron relativistic factor. The reason for such disagreement is as follows: in the case of the positron annihilation with the electron of the separate atom, the positron moves in the Coulomb field of the atom, and so positron wave functions in such field should be used. But in our case, the movement of the positron is controlled by a continuous potential of a system of crystal planes. Therefore, positron wave function in direction perpendicular planes is transverse wave function (solution of the Schrodinger-like equation), while in directions parallel to planes is a plane wave.

 The differential cross-section  $d\sigma^B/d\Omega$ of the positron annihilation with the $K$-shell electron of the separate atom in Born approximation we obtain by the way similar to the one used in this paper by replacing wave function of the channeled positron by a plane wave. For simplicity, we use a hydrogen-like wave function for $K$-shell electrons. As a result, we get
\begin{equation}
  \frac{d\sigma^B}{d\Omega} = \frac{32\pi a_B^3\alpha c(\gamma^2-1) m Z^5\hbar^7\sin^2\Theta}{\gamma[-2 a_B^2c^2(\gamma +1)\sqrt{\gamma^2-1} m^2 \cos\Theta +2 a_B^2 c^2\gamma(\gamma +1) m^2+Z^2 \hbar^2]^4}
 \label{eq31}\: .
\end{equation}
 where $a_B$ is Bohr radius.
\begin{figure}[h]
\centering\noindent
\includegraphics[width=55mm]{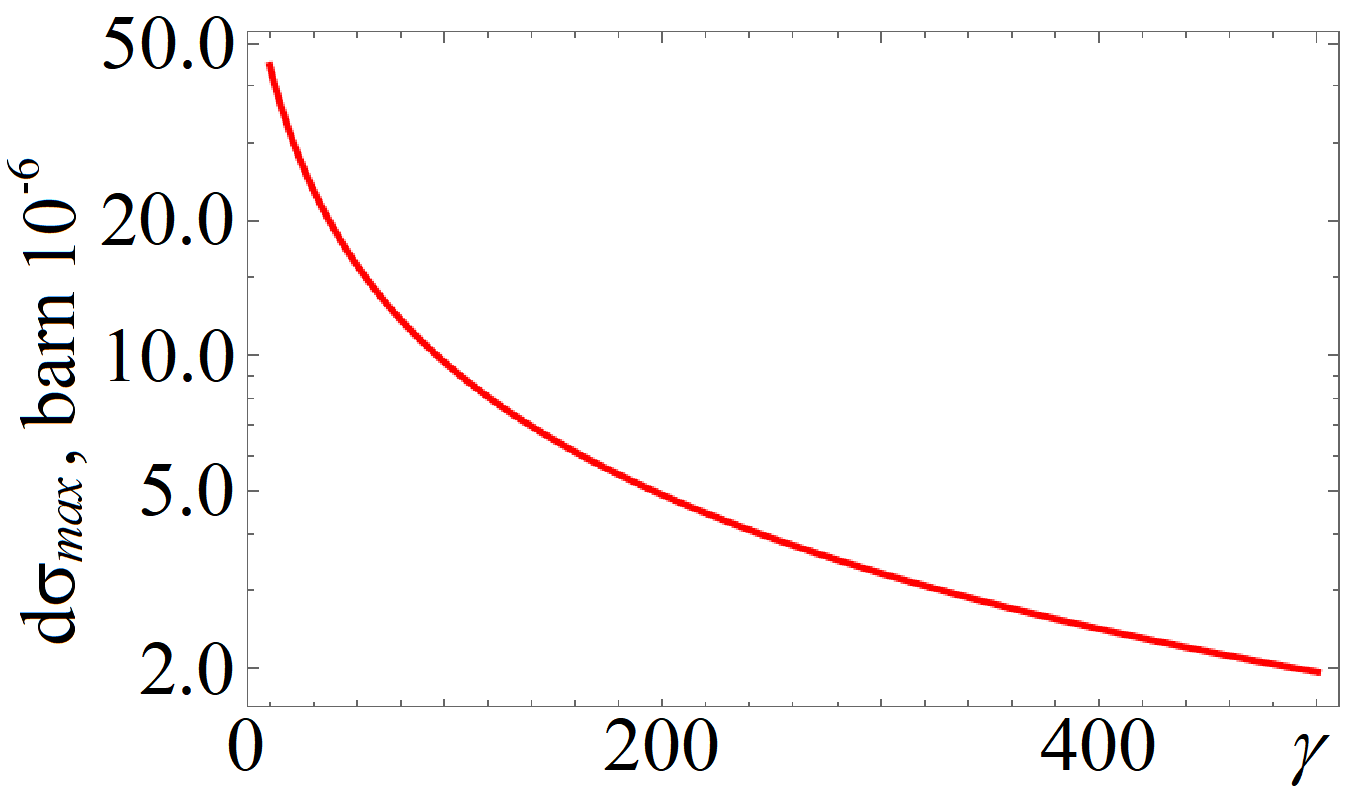}
\caption{Dependences of the maximum value of the cross-section $d\sigma^B_{max}/d\Omega$ of positron SPA on Si atom ($Z = 14$) as a function of the relativistic factor of channeled positrons (a logarithmic scale is used).}
\label{fig9}
\end{figure}

 Result of calculation of the maximum value $d\sigma^B_{max}/d\Omega$ of the differential cross-section $d\sigma^B$ of the positron annihilation with the electron of the separate atom in Born approximation shown in Fig.~\ref{fig9} and gives dependence qualitatively consistent with our results.

\section{Conclusions}
\label{L7}

 In the framework of QED for the first time, is derived differential cross-section of the SPA of positron passing through a crystal in the planar channeled mode with a $K$-shell electron of one of the crystal atoms. Using the obtained formula, we obtained the following results:
\begin{itemize}
  \item The cross-section of SPA of channeled positron for fixed positron transverse energy level depends on the entry angle of positron with respect to the crystal plane:
  \begin{itemize}
    \item For positrons that occupy low levels of transverse energy, the value $d\sigma_{max}/d\Omega$ decreases with increasing angle of entry of the positron into the crystal.
    \item For positrons located at the highest sub-barrier transverse energy band, the picture is the opposite: the magnitude of the maximum value $d\sigma_{max}/d\Omega$ increases with increasing angle of entry of the positron into the crystal.
  \end{itemize}
  \item Neve the less, in the case of SPA of a channeled positron, some kind of orientation dependence should be observed.
\end{itemize}

 In a real experiment, at any angle of incidence (not exceeding the Lindhard angle) of the positron into the crystal, all sub-barrier levels of the transverse energy are populated (with some probability) and therefore, some average curve will be observed.

\section*{Acknowledgements}
\label{L8}

 This research is carried out at the Tomsk Polytechnic University within the framework of the TPU Competitiveness Enhancement Program grant.

\bibliographystyle{SciPost_bibstyle} % Include this style file here only if you are not using our template
%\bibliography{SciPost_Example_BiBTeX_File.bib}

\nolinenumbers

\end{document}